\documentclass[dvips]{article}
\usepackage{icrc2011}
\title{Multiwavelength Modelling of the Globular Cluster Terzan 5}

\newcommand{\etal}{\MakeLowercase{\textit{et al. }}} 
\shorttitle{I. B\"usching \etal Multiwavelength Modelling of Terzan 5}

\authors{I. B\"usching$^{1,2}$, C. Venter$^{2}$, A. Kopp$^{3,2}$, O.C. de Jager$^{2}$, \& A.-C. Clapson$^{4}$}
\afiliations{$^1$Institut f\"ur Theoretische Physik, Lehrstuhl IV: Weltraum- und Astrophysik, Ruhr-Universit\"at Bochum, 44780 Bochum, Germany\\ 
$^2$Centre for Space Research, North-West University, Potchefstroom Campus, Private Bag X6001, Potchefstroom, 2520, South Africa\\
$^3$Institut f\"ur Experimentelle und Angewandte Physik, Christian-Albrechts-Universit\"at zu Kiel, Leibnizstrasse~11, 24118 Kiel, Germany\\
$^4$ Max-Planck-Institut f\"ur Kernphysik, PO Box 103980, 69029 Heidelberg, Germany
}
\email{ib@tp4.rub.de}

\abstract{
Diffuse X-ray emission has recently been detected from the globular cluster (GC) Terzan~5~\cite{Eger10}, extending out to $\sim2.5^\prime$ from the cluster centre~\cite{Clapson11}. This emission may arise from synchrotron radiation (SR) by energetic leptons being injected into the cluster by the resident millisecond pulsar (MSP) population that interact with the cluster field. These leptons may also be reaccelerated in shocks created by collisions of pulsar winds~\cite{BS07}, and may interact with bright starlight and cosmic microwave background (CMB) photons, yielding gamma rays at very high energies (VHE) through the inverse Compton (IC) process. In the GeV range, \textit{Fermi} Large Area Telescope (LAT) has detected a population of GCs~\cite{Abdo10}, very plausibly including Terzan~5, their spectral properties and energetics being consistent with cumulative magnetospheric emission from a population of MSPs. H.E.S.S. has furthermore detected a VHE excess in the direction of Terzan~5~\cite{Abramowski11_Ter5}. One may derive constraints on the number of MSPs, $N_{\rm tot}$, and the radial profiles of the GC B-field, stellar energy density, as well as the diffusion coefficient using the spatially-resolved X-ray, high-energy (HE), and VHE fluxes. If the \textit{Fermi} LAT flux is due to magnetospheric processes, it will scale with the number of visible gamma-ray MSPs, $N_{\rm vis}$. The HE spectrum therefore provides an independent way of constraining the number of MSPs (since $N_{\rm tot}\geq N_{\rm vis}$). Consequently, the synthesis of available multiwavelength data presents a unique opportunity to constrain several parameters of the GC Terzan~5.}


\begin{document}
\maketitle

\section{Introduction}
The first prediction~\cite{BS07} of high-energy (HE) and very-high-energy (VHE) fluxes from globular clusters (GC) invoked inverse Compton (IC) processes of relativistic leptons injected by the embedded population of around~100 millisecond pulsars (MSPs) interacting with the background radiation fields. The MSPs are sources of leptons with energies up to a few TeV, which are accelerated in their magnetospheres by very large electric fields (e.g., \cite{Buesching08}). Additionally, these leptons may be further accelerated in shocks resulting from colliding pulsar winds. Thus,~\cite{BS07} assumed mono-energetic as well as power-law injection spectra (corresponding to the above two acceleration scenarios), and calculated the upscattering of stellar photons (coming from the many old-type stars in the cluster core) and cosmic microwave background (CMB) photons as these energetic particles diffuse outwards. Neglecting synchrotron radiation (SR) losses in their calculation, it was found that GCs should be 
detectable as unpulsed, point-like sources for Cherenkov telescopes, depending of the number of MSPs ($N_{\rm tot}$) in the cluster, as well as the particle efficiency $\eta_e$ (fraction of MSP spin-down power converted into particles). 

The pulsed gamma-ray component coming from particles accelerated inside of the MSP magnetospheres was next calculated~\cite{Venter08}, while a similar calculation to that of~\cite{BS07} of the unpulsed IC component was performed~\cite{Venter09_GC} using an injection spectrum calculated from first principles and which is the result of acceleration and curvature radiation (CR) losses in the MSP magnetospheres, assuming no further particle acceleration. It was found that 47~Tucanae and Terzan~5 may be visible for H.E.S.S., depending on the assumed model parameters, particularly $N_{\rm tot}$ and cluster B-field $B$ (the model fixing $\eta_e$ to $\sim0.007$ due to the magnetospheric acceleration process). 

Recently, H.E.S.S.\ published upper limits on the VHE gamma-ray emission from the GC 47~Tucanae, allowing us to infer $N_{\rm tot}\sim30-40$ for $B\sim10\mu$G, but $N_{\rm tot}$ becoming quite larger for $B<5\,\mu$G or $B>\,30\mu$G. Next, \textit{Fermi} LAT detected HE emission from 47~Tucanae, with the spectrum being consistent with collective pulsed emission from about $50-60$ MSPs in the cluster (\cite{Venter09_GC} inferred $N_{\rm vis}\sim50$). Since then, \textit{Fermi} has detected several GCs~\cite{Abdo09,Abdo10,Kong10,Tam11}, very plausibly including Terzan~5. Following these detections, an IC scenario was considered to explain the HE fluxes seen by \textit{Fermi}~\cite{Cheng10}, as an alternative to the usual CR assumption. For certain parameters, the \textit{Fermi} flux may be reproduced, also predicting spectral components that should be visible in the VHE domain. However, such unpulsed IC components seem less dominant in the case of the cluster NGC~6624, where the HE emission seems to come almost 
exclusively from a single pulsar in the GC, PSR~J1823$-$3021A~\cite{Parent11}. 

H.E.S.S.\ has just announced a VHE excess in the direction of, but offset from the centre of, Terzan~5~\cite{Abramowski11_Ter5}. In addition, diffuse X-ray emission from Terzan~5 was measured~\cite{Eger10}, peaking at its centre but smoothly decreasing outwards, possibly resulting from SR. Also, several radio structures have been identified in the vicinity of Terzan~5, although no reliable estimate could be made for the radio index~\cite{Clapson11}. Lastly, new measurements of Terzan~5's distance~\cite{Ferraro09,Valenti07}, core radius, half-mass radius, tidal radius, and total luminosity~\cite{Lanzoni10} have become available.

In view of all these observational developments and modelling efforts, the aim of this paper is to model the SR flux components expected from Terzan~5 using updated parameters and refined model assumptions. A preliminary IC flux calculation has been presented elsewhere~\cite{Venter11_Fermi}.

\section{The Model}
Some key conclusions have been reached~\cite{BS07} regarding the modelling of GCs which may guide future efforts:
  \begin{itemize}
    \item Pulsar winds mainly interact among themselves, as stellar winds are confined to only a small circumstellar region, as inferred from a simple colliding wind model.
    \item Since the cluster B-field should be less than $B\sim10^{-4}\,$G, and given the typical energy densities of soft photon fields, leptons injected into the cluster mainly lose energy by the IC process, as the latter energy densities exceed the magnetic energy density. SR losses may therefore typically be neglected.
    \item Acceleration of leptons are limited by an advection process along the surface of the shock for typical GC parameters (and not the shock structure, or IC and SR losses), yielding maximum lepton energies of $\sim4-40$~TeV.
    \item IC radiation decreases rapidly when moving out from the GC centre. This should result in point-like gamma-ray sources for Cherenkov telescopes.
  \end{itemize}
In contrast to the last point above, an alternative IC calculation~\cite{Cheng10} solving a cosmic-ray diffusion equation (and using a slightly different stellar photon energy density profile) predicted that most of the HE radiation comes from a region beyond the GC core, implying that GCs should be extended sources. Also, we will include SR losses in our calculation below (for more details, see \cite{Venter08_gamma}).

In this paper, we perform a similar calculation, but using updated structural parameters, a much larger bolometric luminosity, and a distance of $d=$5.9\,kpc \cite{Ferraro09,Lanzoni10,Valenti07} to calculate the radiation and escape losses and resulting unpulsed fluxes, assuming Bohm diffusion and B-fields of $B=1$\,$\mu$G and $B=10$\,$\mu$G. We use an energy density profile similar to that given by~\cite{BS07}, assuming bright starlight and CMB as soft photon targets with central energy densities of several hundred eV/cm$^3$ (for a temperature $T=4~500$\,K) and 0.27\,eV/cm$^3$ (for $T=2.76$\,K). Importantly, we also use a power-law injection spectrum, normalized to the total power output of the particles
\begin{equation}
   L_e = N_{\rm tot}\eta_e\langle\dot{E}_{\rm rot}\rangle,
\end{equation}
assuming $\eta_e=0.01$ and an average MSP spindown of $\langle\dot{E}_{\rm rot}\rangle=2\times10^{34}$~erg s$^{-1}$. This spectral shape will result from  reacceleration of particles in the cluster, after having escaped from the MSP magnetospheres (see Section~\ref{sec:lep}). Since the details of this process are not clear, its actual shape and energy cutoffs will have to be constrained by data.

\section{Propagation of Leptons}  
\label{sec:lep}
For our study, we assume the leptons accelerated by MSPs in the GC core to be reaccelerated at shocks in the CG core, leading to a power-law injection spectrum, and then propagate diffusively. As the MPS provide an almost steady particle source, the distribution function $f$, depending on the spatial distance $r$ from the GC core and energy $E$, satisfies the equation
\begin{eqnarray}
-S&=&\nabla\cdot\left(\kappa(r,E) \nabla f\right)-\frac{\partial }{\partial E}\left(\dot E f\right).
\label{eq:diffusion}
\end{eqnarray}
Here we use the form of the spatial diffusion coefficient,
\begin{eqnarray}
\kappa(r,E)&=&\kappa_0\left(r\right)\left(E/1\,TeV\right)^{\alpha} 
\label{eq:kappa}
\end{eqnarray}
that is inspired by the form of the Galactic diffusion coefficient with $\alpha\,=\,0.6$ in plain diffusion models.
 $\dot E\,={\dot E}(r,E)$ is the rate of lepton energy losses due to IC and SR that depends strongly on the distance from the GC centre due to the spatial variation of the radiation field; $S$ is the source term. For this inital study, we keep $\kappa\,=\,\kappa(E)$. 
As the level of turbulence inside the GC core will be higher than in the surrounding medium, we will allow for a radial dependence of the diffusion coefficient  $\kappa(r)$ in further studies. Eq.~(\ref{eq:diffusion}) is solved numerically.
\section{Results}
\begin{figure*}
\begin{center}
\includegraphics[width=1.5\columnwidth]{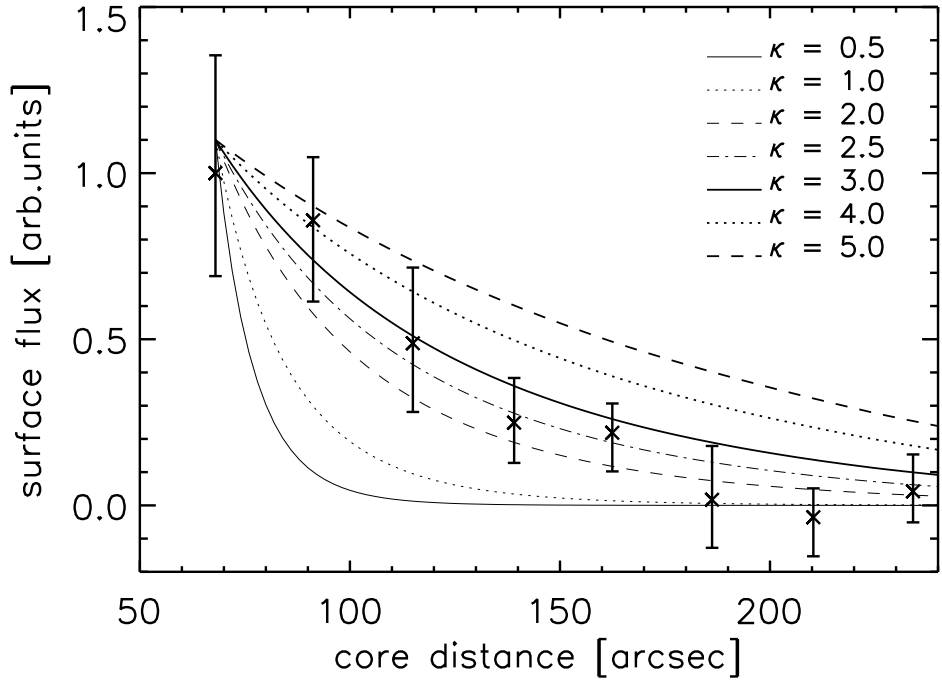}
\end{center}
\caption{Radial profile of the synchrotron flux at 1\,keV for different diffusion coefficients inside the GC. Comparing our calculations to the diffuse X-ray profile measured by~\cite{Eger10}, we find a best-fit value of $\kappa\,=\,$2.5\,kpc$^2$Myr$^{-1}$ at 1\,TeV.}\label{fig1}
\end{figure*}

\begin{figure}
 \includegraphics[height=0.52\columnwidth]{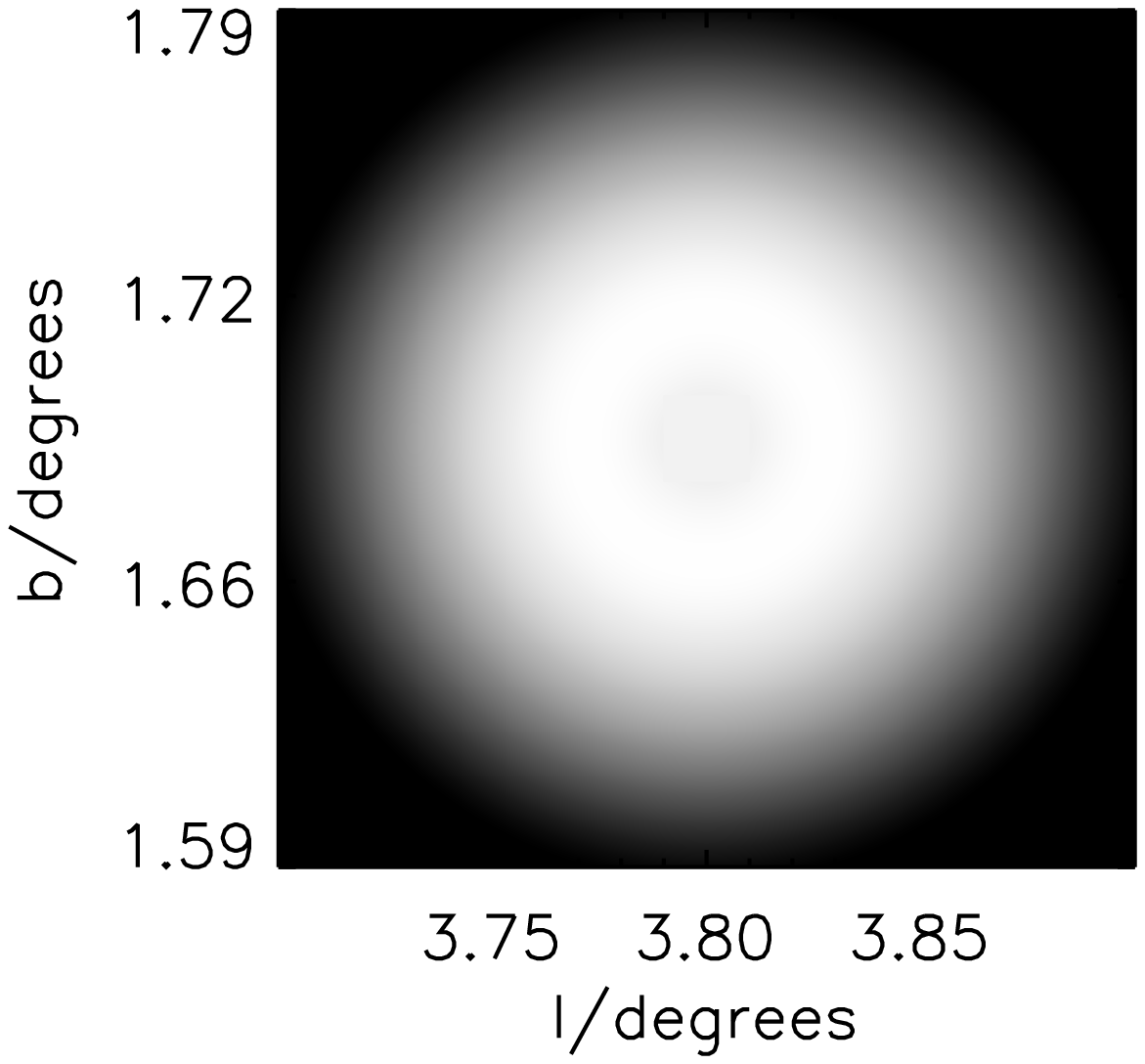}
 \includegraphics[height=0.52\columnwidth]{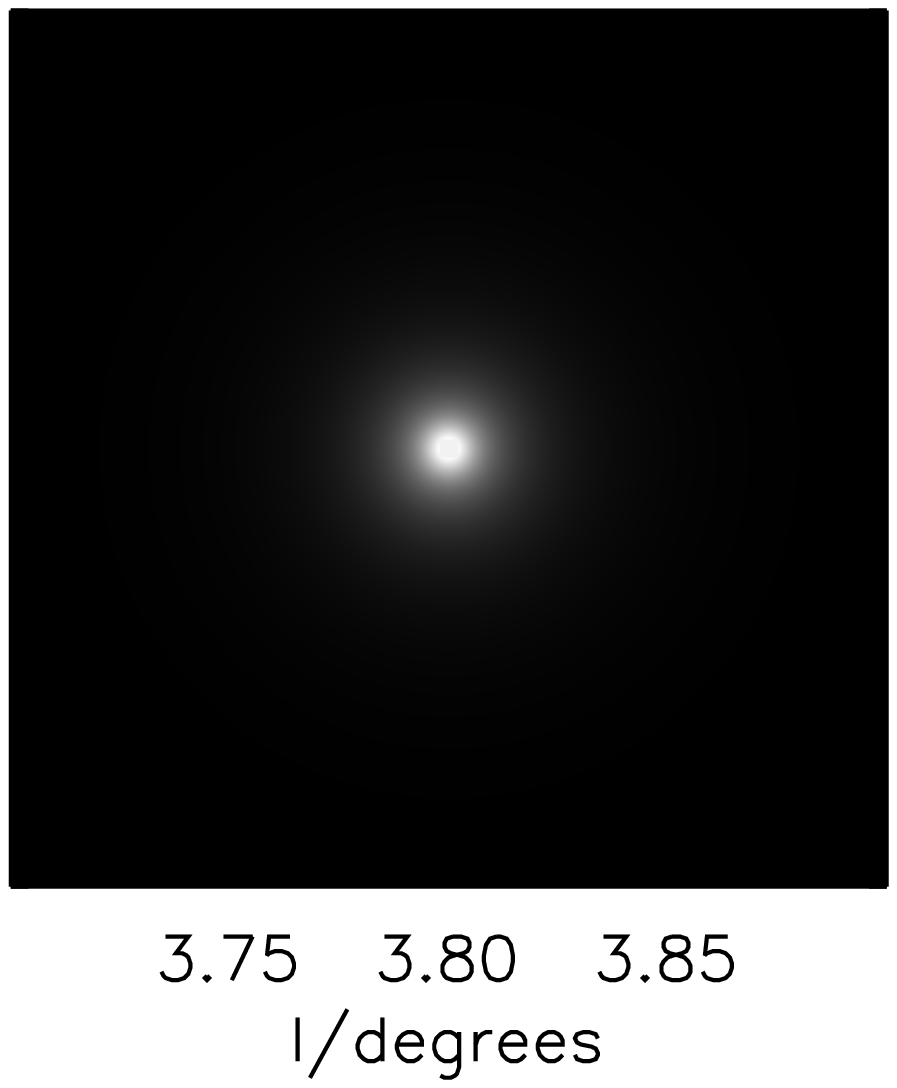}
\caption{Sample sky maps of the synchrotron flux, centred on the GC centre for a diffusion coefficient of 2.5\,kpc$^2$Myr$^{-1}$ at 1\,TeV. The left panel is for 11\,cm, while the right panel is for 1\,keV.
}\label{fig2}
\end{figure}

Firstly, we scaled the pulsed CR component \cite{Venter09_GC,Venter11_Fermi} to fit the \textit{Fermi} LAT data \cite{Abdo10}. This implies a number of visible MSPs $N_{\rm vis}\approx 60\pm 30$. This is also consistent with the estimate of $180^{+90}_{-100}$~\cite{Abdo10}, and formally presents a lower limit for $N_{\rm tot}$, since $N_{\rm tot}\geq N_{\rm vis}$. However, the pair-starved model may overpredict the magnetospheric CR flux by a factor of a few, and furthermore may not be valid for all MSPs. Indeed, light curve modelling \cite{Venter09} has shown that only a small fraction of the current gamma-ray MSP population may be described using the pair-starved model. This argues that $N_{\rm vis}$ may be even larger. For comparison, there are~34 radio-detected MSPs in this cluster\footnote{http://www.naic.edu/$\sim$pfreire/GCpsr.html}, while it is estimated that the total number should be $\sim60-200$~\cite{FG00}.

Using the H.E.S.S.-measured spectrum in conjunction with the HE data, one may next constrain $N_{\rm tot}$ and the cluster $B$-field. The Cherenkov Telescope Array (CTA) may provide more stringent future constraints on these parameters.

We were able to constrain the diffusion coefficient $\kappa$ using the radial profile of the diffuse X-ray emission~\cite{Eger10}. Figure~\ref{fig1} indicates different radial profiles of the SR flux at 1~keV, calculated for different values of $\kappa$, while the associated SR flux sky maps at 11~cm and 1~keV are shown in Figure~\ref{fig2} for the best-fit value of $\kappa\,=\,$2.5\,kpc$^2$Myr$^{-1}$ at 1\,TeV.

\section{Extended, Offset VHE Excess?}
The scenario where MSPs are the main injectors of relativistic particles into the GC implies that the resulting multiwavelength flux due to SR and IC should be centred on the GC if most of the MSPs are located within the core radius. Although this scenario can explain the source energetics (for typical GC parameters), it is intriguing that the VHE excess seen by H.E.S.S.~\cite{Abramowski11_Ter5} appears to be extended and offset from the GC centre. There may be a few reasons for this:
\begin{itemize}
  \item There may be radio-faint and / or off-beam MSPs outside of the GC core that inject leptons, but have not been observed to produce pulsed emission.
  \item A bulk transport of relativistic particles to another acceleration site may have occurred.
  \item The VHE excess may be due to a see-through background source.
  \item Some other sources inside the GC beside MSPs may be responsible for / contribute to the VHE flux.
\end{itemize}
In the first case, it is expected that MSPs have very wide radio and gamma-ray beams~(e.g.,~\cite{Venter09}), so they will only be invisible if they are intrinsically faint. Also, the cumulative pulsed emission consisting of single-MSP emission pulsed at different periods, as well as the typical short MSP periods, complicate blind searches for new ones. Second, a transport of particles is expected to leave a trail of radiation, although identification of such an emission structure may be limited by the angular resolution of the telescope in the case of gamma rays. Third, it is quite improbable that the VHE emission will coincide with Terzan~5 by chance, but not impossible. Lastly, detailed modelling will be needed to demonstrate that the predicted spectral normalization and shape correctly reproduce the data in case other sources of HE leptons are invoked, as the MSP scenario has been reasonably successful at explaining the pulsed spectra of GCs. These tentative ideas need to be developed further to assess 
their viability. It should also be remembered that Terzan~5 may be considered a non-typical GC, showing signs of a complex stellar formation history~\cite{Ferraro09}.

\section{Conclusions}
One may derive important constraints on the GC Terzan~5 by using multiwavelength data. The combination of HE and VHE data will already constrain $N_{\rm tot}$ and $B$, for a given energy density and diffusion coefficient, using fluxes averaged over the cluster's extent. However, a more refined model including particle transport and radiation as a function of radius will allow the prediction of SR and ICS flux profiles, in addition to the total fluxes.

Modelling the CR component~\cite{Venter11_Fermi}, we found that $N_{\rm vis}\approx 60\pm 30$, consistent with the estimates presented in~\cite{FG00,Abdo10}, and representing a lower limit for $N_{\rm tot}$.
 
Assuming that the electrons injected by the embedded MSPs are reaccelerated at shocks in the GC core, and using a diffusion coefficient with the same energy dependence as in models of Galactic CR propagation, we find that a diffusion coefficient of 2.5\,kpc$^2$Myr$^{-1}$ at 1\,TeV describes the radial profile of the X-ray flux as reported by \cite{Eger10} best (assuming a cluster field of $B=10\mu$G).

Future work will include attempts to constrain the radial profile of the diffusion coefficient, energy density, and cluster field as well as the number of host MSPs and the particle injection spectra using radio, X-ray, and gamma-ray data.


This research is based on work supported by the South African National Research Foundation.

\clearpage

\end{document}